# Lognormal and Gamma Mixed Negative Binomial Regression

Mingyuan Zhou  MZ1@EE.DUKE.EDU
Lingbo Li  LL83@DUKE.EDU
David Dunson  DUNSON@STAT.DUKE.EDU
Lawrence Carin  LCARIN@EE.DUKE.EDU
Duke University, Durham NC 27708, USA

## Abstract

In regression analysis of counts, a lack of simple and efficient algorithms for posterior computation has made Bayesian approaches appear unattractive and thus underdeveloped. We propose a lognormal and gamma mixed negative binomial (NB) regression model for counts, and present efficient closed-form Bayesian inference; unlike conventional Poisson models, the proposed approach has two free parameters to include two different kinds of random effects, and allows the incorporation of prior information, such as sparsity in the regression coefficients. By placing a gamma distribution prior on the NB dispersion parameter $r$, and connecting a lognormal distribution prior with the logit of the NB probability parameter $p$, efficient Gibbs sampling and variational Bayes inference are both developed. The closed-form updates are obtained by exploiting conditional conjugacy via both a compound Poisson representation and a Polya-Gamma distribution based data augmentation approach. The proposed Bayesian inference can be implemented routinely, while being easily generalizable to more complex settings involving multivariate dependence structures. The algorithms are illustrated using real examples.

## 1. Introduction

In numerous scientific studies, the response variable is a count $y = 0, 1, 2, \cdots$, which we wish to explain with a set of covariates $\boldsymbol{x} = [1, x_1, \cdots, x_P]^T$ as $\mathbb{E}[y|\boldsymbol{x}] = g^{-1}(\boldsymbol{x}^T\boldsymbol{\beta})$, where $\boldsymbol{\beta} = [\beta_0, \cdots, \beta_P]^T$ are the regression coefficients and $g$ is the canonical link function in generalized linear models (GLMs) (McCullagh & Nelder, 1989; Long, 1997; Cameron & Trivedi, 1998; Agresti, 2002; Winkelmann, 2008). Regression models for counts are usually nonlinear and have to take into consideration the specific properties of counts, including discreteness and nonnegativity, and often characterized by overdispersion (variance greater than the mean). In addition, we may wish to impose a sparse prior in the regression coefficients for counts, which is demonstrated to be beneficial for regression analysis of both Gaussian and binary data (Tipping, 2001).

Count data are commonly modeled with the Poisson distribution $y \sim \text{Pois}(\lambda)$, whose mean and variance are both equal to $\lambda$. Due to heterogeneity (difference between individuals) and contagion (dependence between the occurrence of events), the varance is often much larger than the mean, making the Poisson assumption restrictive. By placing a gamma distribution prior with shape $r$ and scale $p/(1-p)$ on $\lambda$, a negative binomial (NB) distribution $y \sim \text{NB}(r, p)$ can be generated as $f_Y(y) = \int_0^\infty \text{Pois}(y; \lambda)\text{Gamma}\left(\lambda; r, \frac{p}{1-p}\right) d\lambda = \frac{\Gamma(r+y)}{y!\Gamma(r)}(1-p)^r p^y$, where $\Gamma(\cdot)$ denotes the gamma function, $r$ is the nonnegative dispersion parameter and $p$ is a probability parameter. Therefore, the NB distribution is also known as the gamma-Poisson distribution. It has a variance $rp/(1-p)^2$ larger than the mean $rp/(1-p)$, and thus it is usually favored over the Poisson distribution for modeling overdispersed counts.

The regression analysis of counts is commonly performed under the Poisson or NB likelihoods, whose parameters are usually estimated by finding the maximum of the nonlinear log likelihood (Long, 1997; Cameron & Trivedi, 1998; Agresti, 2002; Winkelmann, 2008). The maximum likelihood estimator (MLE), however, only provides a point estimate and does not allow the incorporation of prior information, such as sparsity in the regression coefficients. In addition, the MLE of the NB dispersion parameter $r$ often lacks robustness and may be severely biased or even fail to converge if the sample size is small, the mean is small or if $r$ is large (Saha & Paul, 2005; Lloyd-Smith, 2007).

Compared to the MLE, Bayesian approaches are able to model the uncertainty of estimation and to incorporate prior information. In regression analysis of counts, however, the lack of simple and efficient algorithms for posterior computation has seriously limited routine applications of Bayesian approaches, making Bayesian analysis of counts appear unattractive and thus underdeveloped. For instance, for the NB dispersion parameter $r$, the only available closed-form Bayesian solution relies on approximating the ratio of two gamma functions using a polynomial expansion (Bradlow et al., 2002); and for the regression coefficients $\boldsymbol{\beta}$, Bayesian solutions usually involve computationally intensive Metropolis-Hastings algorithms, since the conjugate prior for $\boldsymbol{\beta}$ is not known under the Poisson and NB likelihoods (Chib et al., 1998; Chib & Winkelmann, 2001; Winkelmann, 2008).

In this paper we propose a lognormal and gamma mixed NB regression model for counts, with default Bayesian analysis presented based on two novel data augmentation approaches. Specifically, we show that the gamma distribution is the conjugate prior to the NB dispersion parameter $r$, under the compound Poisson representation, with efficient Gibbs sampling and variational Bayes (VB) inference derived by exploiting conditional conjugacy. Further we show that a lognormal prior can be connected to the logit of the NB probability parameter $p$, with efficient Gibbs sampling and VB inference developed for the regression coefficients $\boldsymbol{\beta}$ and the lognormal variance parameter $\sigma^2$, by generalizing a Polya-Gamma distribution based data augmentation approach in Polson & Scott (2011). The proposed Bayesian inference can be implemented routinely, while being easily generalizable to more complex settings involving multivariate dependence structures. We illustrate the algorithm with real examples on univariate count analysis and count regression, and demonstrate the advantages of the proposed Bayesian approaches over conventional count models.

## 2. Regression Models for Counts

The most basic regression model for counts is the Poisson regression model (Long, 1997; Cameron & Trivedi, 1998; Winkelmann, 2008), which can be expressed as

$$y_i \sim \text{Pois}(\lambda_i), \quad \lambda_i = \exp(\boldsymbol{x}_i^T \boldsymbol{\beta}) \quad (1)$$

where $\boldsymbol{x}_i = [1, x_{i1}, \cdots, x_{iP}]^T$ is the covariate vector for sample $i$. The Newton-Raphson method can be used to iteratively find the MLE of $\boldsymbol{\beta}$ (Long, 1997). A serious constraint of the Poisson regression model is that it assumes equal-dispersion, i.e., $\mathbb{E}[y_i|\boldsymbol{x}_i] = \text{Var}[y_i|\boldsymbol{x}_i] = \exp(\boldsymbol{x}_i^T \boldsymbol{\beta})$. In practice, however, count data are often overdispersed, due to heterogeneity and contagion (Winkelmann, 2008). To model overdispersed counts, the Poisson regression model can be modified as

$$y_i \sim \text{Pois}(\lambda_i), \quad \lambda_i = \exp(\boldsymbol{x}_i^T \boldsymbol{\beta}) \epsilon_i \quad (2)$$

where $\epsilon_i$ is a nonnegative multiplicative random-effect term to model individual heterogeneity (Winkelmann, 2008). Using both the law of total expectation and the law of total variance, it can be shown that

$$\mathbb{E}[y_i|\boldsymbol{x}_i] = \exp(\boldsymbol{x}_i^T \boldsymbol{\beta}) \mathbb{E}[\epsilon_i] \quad (3)$$

$$\text{Var}[y_i|\boldsymbol{x}_i] = \mathbb{E}[y_i|\boldsymbol{x}_i] + \frac{\text{Var}[\epsilon_i]}{\mathbb{E}^2[\epsilon_i]} \mathbb{E}^2[y_i|\boldsymbol{x}_i]. \quad (4)$$

Thus $\text{Var}[y_i|\boldsymbol{x}_i] \geq \mathbb{E}[y_i|\boldsymbol{x}_i]$ and we obtain a regression model for overdispersed counts. We show below that both the gamma and lognormal distributions can be used as the nonnegative prior on $\epsilon_i$.

### 2.1. The Negative Binomial Regression Model

The NB regression model (Long, 1997; Cameron & Trivedi, 1998; Winkelmann, 2008; Hilbe, 2007) is constructed by placing a gamma prior on $\epsilon_i$ as

$$\epsilon_i \sim \text{Gamma}(r, 1/r) = \frac{r^r}{\Gamma(r)} \epsilon_i^{r-1} e^{-r\epsilon_i} \quad (5)$$

where $\mathbb{E}[\epsilon_i] = 1$ and $\text{Var}[\epsilon_i] = r^{-1}$. Marginalizing out $\epsilon_i$ in (2), we have a NB distribution parameterized by mean $\mu_i = \exp(\boldsymbol{x}_i^T \boldsymbol{\beta})$ and inverse dispersion parameter $\phi$ (the reciprocal of $r$) as $f_Y(y_i) = \frac{\Gamma(\phi^{-1}+y_i)}{y_i!\Gamma(\phi^{-1})} \left(\frac{\phi^{-1}}{\phi^{-1}+\mu_i}\right)^{\phi^{-1}} \left(\frac{\mu_i}{\phi^{-1}+\mu_i}\right)^{y_i}$, thus

$$\mathbb{E}[y_i|\boldsymbol{x}_i] = \exp(\boldsymbol{x}_i^T \boldsymbol{\beta}) \quad (6)$$

$$\text{Var}[y_i|\boldsymbol{x}_i] = \mathbb{E}[y_i|\boldsymbol{x}_i] + \phi \mathbb{E}^2[y_i|\boldsymbol{x}_i]. \quad (7)$$

The MLEs of $\boldsymbol{\beta}$ and $\phi$ can be found numerically with the Newton-Raphson method (Lawless, 1987).

### 2.2. The Lognormal-Poisson Regression Model

A lognormal-Poisson regression model (Breslow, 1984; Long, 1997; Agresti, 2002; Winkelmann, 2008) can be constructed by placing a lognormal prior on $\epsilon_i$ as

$$\epsilon_i \sim \ln \mathcal{N}(0, \sigma^2) \quad (8)$$

where $\mathbb{E}[\epsilon_i] = e^{\sigma^2/2}$ and $\text{Var}[\epsilon_i] = e^{\sigma^2}\left(e^{\sigma^2} - 1\right)$. Using (3) and (4), we have

$$\mathbb{E}[y_i|\boldsymbol{x}_i] = \exp(\boldsymbol{x}_i^T \boldsymbol{\beta} + \sigma^2/2) \quad (9)$$

$$\text{Var}[y_i|\boldsymbol{x}_i] = \mathbb{E}[y_i|\boldsymbol{x}_i] + \left(e^{\sigma^2} - 1\right) \mathbb{E}^2[y_i|\boldsymbol{x}_i]. \quad (10)$$

Compared to the NB model, there is no analytical form for the distribution of $y_i$ if $\epsilon_i$ is marginalized out and the MLE is less straightforward to calculate, making it less commonly used. However, Winkelmann (2008) suggests to reevaluate the lognormal-Poisson model, since it is appealing in theory and may fit the data better. The inverse Gaussian distribution prior can also be placed on $\epsilon_i$ to construct a heavier-tailed alternative to the NB model (Dean et al., 1989), whose density functions are shown to be virtually identical to the lognormal-Poisson model (Winkelmann, 2008).

## 3. The Lognormal and Gamma Mixed Negative Binomial Regression Model

To explicitly model the uncertainty of estimation and incorporate prior information, Bayesian approaches appear attractive. Bayesian analysis of counts, however, is seriously limited by the lack of efficient inference, as the conjugate prior for the regression coefficients $\boldsymbol{\beta}$ is unknown under the Poisson and NB likelihoods (Winkelmann, 2008), and the conjugate prior for the NB dispersion parameter $r$ is also unknown.

To address these issues, we propose a lognormal and gamma mixed NB regression model for counts, termed here the LGNB model, where a lognormal prior $\ln \mathcal{N}(0, \sigma^2)$ is placed on the multiplicative random effect term $\epsilon_i$ and a gamma prior is placed on $r$. Denoting $p_i = \frac{e^{\psi_i}}{1+e^{\psi_i}} = \frac{\exp(\boldsymbol{x}_i^T \boldsymbol{\beta}) \epsilon_i}{1+\exp(\boldsymbol{x}_i^T \boldsymbol{\beta}) \epsilon_i}$, and $\operatorname{logit}(p_i) = \ln \frac{p_i}{1-p_i}$, the LGNB model is constructed as

$$y_i \sim \operatorname{NB}(r, p_i), \quad \psi_i = \operatorname{logit}(p_i) = \boldsymbol{x}_i^T \boldsymbol{\beta} + \ln \epsilon_i \quad (11)$$

$$\epsilon_i \sim \ln \mathcal{N}(0, \varphi^{-1}), \qquad \varphi \sim \operatorname{Gamma}(e_0, 1/f_0) \quad (12)$$

$$\boldsymbol{\beta} \sim \prod_{p=0}^{P} \mathcal{N}(0, \alpha_p^{-1}), \qquad \alpha_p \sim \operatorname{Gamma}(c_0, 1/d_0) \quad (13)$$

$$r \sim \operatorname{Gamma}(a_0, 1/h), \quad h \sim \operatorname{Gamma}(b_0, 1/g_0) \quad (14)$$

where $\varphi = \sigma^{-2}$ and $a_0$, $b_0$, $c_0$, $d_0$, $e_0$, $f_0$ and $g_0$ are gamma hyperparameters (they are set as 0.01 in experiments). Since $y_i \sim \operatorname{NB}(r, p_i)$ in (11) can be augmented into a gamma-Poisson structure as $y_i \sim \operatorname{Pois}(\lambda_i)$, $\lambda_i \sim \operatorname{Gamma}(r, \exp(\boldsymbol{x}_i^T \boldsymbol{\beta}) \epsilon_i)$, the LGNB model can also be considered as a lognormal-gamma-gamma-Poisson regression model. Denoting $\mathbf{X} = [\boldsymbol{x}_1^T, \cdots, \boldsymbol{x}_N^T]^T$, we may equivalently express $\boldsymbol{\psi} = [\psi_1, \cdots, \psi_N]^T$ in the above model as

$$\boldsymbol{\psi} \sim \mathcal{N}(\mathbf{X}\boldsymbol{\beta}, \varphi^{-1}\mathbf{I}). \quad (15)$$

If we marginalize out $h$ in (14), we obtain a beta prime distribution prior $r \sim \beta'(a_0, b_0, 1, g_0)$. If we marginalize out $\alpha_p$ in (13), we obtain a Student-t prior for $\beta_p$, the sparsity-promoting prior used in Tipping (2001); Bishop & Tipping (2000) for regression analysis of both Gaussian and binary data. Note that $\boldsymbol{\beta}$ is connected to $p_i$ with a logit link, which is key to deriving efficient Bayesian inference.

### 3.1. Model Properties and Model Comparison

Using the laws of total expectation and total variance and the moments of the NB distribution, we have

$$\mathbb{E}[y_i | \boldsymbol{x}_i] = \mathbb{E}_{\epsilon_i}[\mathbb{E}[y_i | \boldsymbol{x}_i, \epsilon_i]] = \exp(\boldsymbol{x}_i^T \boldsymbol{\beta} + \sigma^2/2 + \ln r) \quad (16)$$

$$\operatorname{Var}[y_i | \boldsymbol{x}_i] = \mathbb{E}_{\epsilon_i}[\operatorname{Var}[y_i | \boldsymbol{x}_i, \epsilon_i]] + \operatorname{Var}_{\epsilon_i}[\mathbb{E}[y_i | \boldsymbol{x}_i, \epsilon_i]]$$
$$= \mathbb{E}[y_i | \boldsymbol{x}_i] + \left(e^{\sigma^2}(1+r^{-1}) - 1\right) \mathbb{E}^2[y_i | \boldsymbol{x}_i]. \quad (17)$$

We define the quasi-dispersion $\kappa$ as the coefficient associated with the mean quadratic term in the variance. As shown in (7) and (10), $\kappa = \phi$ in the NB model and $\kappa = \left(e^{\sigma^2} - 1\right)$ in the lognormal-Poisson model. Apparently, they have different distribution assumptions on dispersion, yet there is no clear evidence to favor one over the other in terms of goodness of fit. In the proposed LGNB model, there are two free parameters $r$ and $\sigma^2$ to adjust both the mean in (16) and dispersion $\kappa = \left(e^{\sigma^2}(1+r^{-1}) - 1\right)$, which become the same as those of the NB model when $\sigma^2 = 0$, and the same as those of the lognormal-Poisson model when $\phi = r^{-1} = 0$. Thus the LGNB model has one extra degree of freedom to incorporate both kinds of random effects, with their proportion automatically inferred.

## 4. Default Bayesian Analysis Using Data Augmentation

As discussed in Section 3, the LGNB model has an advantage of having two free parameters to incorporate both kinds of random effects. We show below that it has an additional advantage in that default Bayesian analysis can be performed with two novel data augmentation approaches, with closed-form solutions and analytical update equations available for both Gibbs sampling and VB inference. One augmentation approach concerns the inference of the NB dispersion parameter $r$ using the compound Poisson representation, and the other concerns the inference of the regression coefficients $\boldsymbol{\beta}$ using the Polya-Gamma distribution.

### 4.1. Inferring the Dispersion Parameter Under the Compound Poisson Representation

We first focus on inference of the NB dispersion parameter $r$ and assume we know $\{p_i\}_{i=1,N}$ and $h$, neglecting the remaining part of the LGNB model at this moment. We comment here that the novel Bayesian inference developed here can be applied to any other scenarios where the conditional posterior of $r$ is proportional to $\prod_{i=1}^{N} \operatorname{NB}(y_i; r, p_i) \operatorname{Gamma}(r; a_0, 1/h)$, for which a hybrid Monte Carlo and a Metropolis-Hastings algorithms had been developed in Williamson et al. (2010) and Zhou et al. (2012), but VB solutions were not yet developed.

As proved in Quenouille (1949), $y \sim \text{NB}(r, p)$ can also be generated from a compound Poisson distribution as

$$y = \sum_{\ell=1}^{L} u_\ell, \quad L \sim \text{Pois}(-r\ln(1-p)), \quad u_\ell \stackrel{iid}{\sim} \text{Log}(p) \quad (18)$$

where $\text{Log}(p)$ corresponds to the logarithmic distribution (Barndorff-Nielsen et al., 2010) with $f_U(k) = -p^k/[k\ln(1-p)]$, $k \in \{1, 2, \ldots\}$, whose probability-generating function (PGF) is

$$G_U(z) = \ln(1-pz)/\ln(1-p), \quad |z| < p^{-1}. \quad (19)$$

Using the conjugacy between the gamma and Poisson distributions, it is evident that the gamma distribution is the conjugate prior for $r$ under this augmentation.

4.1.1. GIBBS SAMPLING FOR $r$

Recalling (18), $y_i \sim \text{NB}(r, p_i)$ can also be generated from the random sum $y_i = \sum_{\ell=1}^{L_i} u_{i\ell}$ with

$$u_{i\ell} \stackrel{iid}{\sim} \text{Log}(p_i), \quad L_i \sim \text{Pois}(-r\ln(1-p_i)). \quad (20)$$

Exploiting conjugacy between (14) and (20), given $L_i$, we have the conditional posterior of $r$ as

$$(r|-) \sim \text{Gamma}\left(a_0 + \sum_{i=1}^{N} L_i, \frac{1}{h - \sum_{i=1}^{N} \ln(1-p_i)}\right) \quad (21)$$

where here and below expressions like $(r|-)$ correspond to random variable (RV) $r$, conditioned on all other RVs. The remaining challenge is finding the conditional posterior of $L_i$. Denote $w_{ij} = \sum_{\ell=1}^{j} u_{i\ell}$, $j = 1, \cdots, y_i$. Since $w_{ij}$ is the summation of $j$ iid $\text{Log}(p_i)$ distributed RVs, using (19), the PGF of $w_{ij}$ is

$$G_{W_{ij}}(z) = [\ln(1-p_iz)/\ln(1-p_i)]^j, \quad |z| < p_i^{-1}.$$

Therefore, we have $L_i \equiv 0$ if $y_i = 0$ and for $1 \leq j \leq y_i$

$$\Pr(L_i = j|-) \propto \Pr(w_{ij} = y_i)\text{Pois}(j; -r\ln(1-p_i))$$
$$= \frac{G_{W_{ij}}^{(y_i)}(0)}{y_i!}\text{Pois}(j; -r\ln(1-p_i))$$
$$= \left.\frac{d^{y_i}}{dz^{y_i}}f_i^j(z)\right|_{z=0}\frac{r^j}{j!y_i!}\exp(r\ln(1-p_i))$$
$$= F(y_i, j)r^j p_i^{y_i}\exp(r\ln(1-p_i)) \quad (22)$$

where $f_i(z) = -\ln(1-p_iz)$ and $\mathbf{F}$ is a lower triangular matrix with $F(1,1) = 1$, $F(m,j) = 0$ if $j > m$, and

$$F(m,j) = \frac{m-1}{m}F(m-1,j) + \frac{1}{m}F(m-1,j-1) \quad (23)$$

if $1 \leq j \leq m$. Using (22), we have

$$\Pr(L_i = j|-) = R_r(y_i, j), \quad j = 0, \cdots, y_i. \quad (24)$$

where $R_r(0,0) = 1$ and

$$R_r(m,j) = F(m,j)r^j \Big/ \sum_{j'=1}^{m} F(m,j')r^{j'}. \quad (25)$$

The values of $\mathbf{F}$ can be iteratively calculated and each row sums to one, e.g., the 4th and 5th rows of $\mathbf{F}$ are

$$\begin{pmatrix} 6/4! & 11/4! & 6/4! & 1/4! & 0 & 0 & \cdots \\ 24/5! & 50/5! & 35/5! & 10/5! & 1/5! & 0 & \cdots \end{pmatrix}.$$

Note that to obtain (22), we use the relationship proved in Lemma 1 of the supplementary material that

$$\left.\frac{1}{m!}\frac{d^m}{dz^m}f_i^j(z)\right|_{z=0} = F(m,j)j!p_i^m, \quad 1 \leq j \leq m. \quad (26)$$

Gibbs sampling for $r$ proceeds by alternately sampling (24) and (21). Note that to ensure numerical stability when $r > 1$, instead of using (25), we may iteratively calculate $\mathbf{R}_r$ in the way we calculate $\mathbf{F}$ in (23). We show in Figure 1 of the supplementary material the matrices $\mathbf{R}_r$ for $r = .1, 1, 10$ and $100$.

4.1.2. VARIATIONAL BAYES INFERENCE FOR $r$

Using VB inference (Bishop & Tipping, 2000; Beal, 2003), we approximate the posterior $p(r, \mathbf{L}|\mathbf{X})$ with $Q(r, \mathbf{L}) = Q_r(r)\prod_{i=1}^{N} Q_{L_i}(L_i)$, and we have

$$Q_{L_i}(L_i) = \sum_{j=0}^{y_i} R_{\tilde{r}}(y_i, j)\delta_j \quad (27)$$

$$Q_r(r) = \text{Gamma}(\tilde{a}, 1/\tilde{h}). \quad (28)$$

where $\langle x \rangle = \mathbb{E}[x]$, $\tilde{r} = \exp(\langle \ln r \rangle)$, $\psi(x)$ is the digamma function, and

$$\langle \ln r \rangle = \psi(\tilde{a}) - \ln \tilde{h}, \quad \langle L_i \rangle = \sum_{j=1}^{y_i} R_{\tilde{r}}(y_i, j)j \quad (29)$$

$$\tilde{a} = a_0 + \sum_{i=1}^{N} \langle L_i \rangle, \quad \tilde{h} = h - \sum_{i=1}^{N} \langle \ln(1-p_i) \rangle. \quad (30)$$

Equations (29)-(30) constitute the VB inference for the NB dispersion parameter $r$, with $\langle r \rangle = \tilde{a}/\tilde{h}$.

### 4.2. Inferring the Regression Coefficients Using the Polya-Gamma Distribution

Denote $\omega_i$ as a random variable drawn from the Polya-Gamma (PG) distribution (Polson & Scott, 2011) as

$$\omega_i \sim \text{PG}(y_i + r, 0). \quad (31)$$

We have $\mathbb{E}_{\omega_i}\left[\exp(-\omega_i\psi_i^2/2)\right] = \cosh^{-(y_i+r)}(\psi_i/2)$. Thus the likelihood of $\psi_i$ in (11) can be expressed as

$$\mathcal{L}(\psi_i) \propto \frac{(e^{\psi_i})^{y_i}}{(1+e^{\psi_i})^{y_i+r}} = \frac{2^{-(y_i+r)}\exp(\frac{y_i-r}{2}\psi_i)}{\cosh^{y_i+r}(\psi_i/2)}$$
$$\propto \exp\left(\frac{y_i-r}{2}\psi_i\right)\mathbb{E}_{\omega_i}\left[\exp(-\omega_i\psi_i^2/2)\right]. \quad (32)$$

Given the values of $\{\omega_i\}_{i=1,N}$ and the prior in (15), the conditional posterior of $\boldsymbol{\psi}$ can be expressed as

$$(\boldsymbol{\psi}|-) \propto \mathcal{N}(\boldsymbol{\psi}; \mathbf{X}\boldsymbol{\beta}, \varphi^{-1}\mathbf{I}) \prod_{i=1}^{N} e^{-\frac{\omega_i}{2}\left(\psi_i - \frac{y_i - r}{2\omega_i}\right)^2} \quad (33)$$

and given the values of $\boldsymbol{\psi}$ and the prior in (31), the conditional posterior of $\omega_i$ can be expressed as

$$(\omega_i|-) \propto \exp(-\omega_i \psi_i^2/2) \text{PG}(\omega_i; y_i + r, 0). \quad (34)$$

### 4.3. Gibbs Sampling Inference

Exploiting conditional conjugacy and the exponential tilting of the PG distribution in Polson & Scott (2011), we can sample in closed-form all latent parameters of the LGNB model described from (11) to (14) as

$$\text{Sampling } L_i \text{ with (24), Sampling } r \text{ with (21)} \quad (35)$$
$$(\omega_i|-) \sim \text{PG}(y_i + r, \psi_i) \quad (36)$$
$$(\boldsymbol{\psi}|-) \sim \mathcal{N}(\boldsymbol{\mu}, \boldsymbol{\Sigma}), \quad (\boldsymbol{\beta}|-) \sim \mathcal{N}(\boldsymbol{\mu}_{\boldsymbol{\beta}}, \boldsymbol{\Sigma}_{\boldsymbol{\beta}}) \quad (37)$$
$$(h|-) \sim \text{Gamma}\left(a_0 + b_0, 1/(g_0 + r)\right) \quad (38)$$
$$(\varphi|-) \sim \text{Gamma}\left(e_0 + \frac{N}{2}, \frac{1}{f_0 + \|\boldsymbol{\psi} - \mathbf{X}\boldsymbol{\beta}\|_2^2/2}\right) \quad (39)$$
$$(\alpha_p|-) \sim \text{Gamma}\left(c_0 + 1/2, 1/(d_0 + \beta_p^2/2)\right) \quad (40)$$

where $\boldsymbol{\Omega} = \text{diag}(\omega_1, \cdots, \omega_N)$, $\mathbf{A} = \text{diag}(\alpha_0, \cdots, \alpha_P)$, $\boldsymbol{y} = [y_1, \cdots, y_N]^T$, $\boldsymbol{\Sigma} = (\varphi\mathbf{I} + \boldsymbol{\Omega})^{-1}$, $\boldsymbol{\mu} = \boldsymbol{\Sigma}[(\boldsymbol{y} - r)/2 + \varphi\mathbf{X}\boldsymbol{\beta}]$, $\boldsymbol{\Sigma}_{\boldsymbol{\beta}} = (\varphi\mathbf{X}^T\mathbf{X} + \mathbf{A})^{-1}$ and $\boldsymbol{\mu}_{\boldsymbol{\beta}} = \varphi\boldsymbol{\Sigma}_{\boldsymbol{\beta}}\mathbf{X}^T\boldsymbol{\psi}$. Note that a PG distributed random variable can be generated from an infinite sum of weighted iid gamma random variables (Devroye, 2009; Polson & Scott, 2011). We provide in the supplementary material a method for accurately truncating the infinite sum.

### 4.4. Variational Bayes Inference

Using VB inference (Bishop & Tipping, 2000; Beal, 2003), we approximate the posterior distribution with $Q = Q_{\boldsymbol{\psi}}(\boldsymbol{\psi})Q_{\boldsymbol{\beta}}(\boldsymbol{\beta})Q_r(r)Q_h(h)Q_\varphi(\varphi)\prod_{p=0}^{P} Q_{\alpha_p}(\alpha_p)$ $\prod_{i=1}^{N}[Q_{L_i}(L_i)Q_{\omega_i}(\omega_i)]$. To exploit conjugacy, defining $Q_{L_i}(L_i)$ as in (27), $Q_r(r)$ as in (28), $Q_{\omega_i}(\omega_i) = \text{PG}(\gamma_{i1}, \gamma_{i2})$, $Q_{\boldsymbol{\psi}}(\boldsymbol{\psi}) = \mathcal{N}(\tilde{\boldsymbol{\mu}}, \tilde{\boldsymbol{\Sigma}})$, $Q_{\boldsymbol{\beta}}(\boldsymbol{\beta}) = \mathcal{N}(\tilde{\boldsymbol{\mu}}_{\boldsymbol{\beta}}, \tilde{\boldsymbol{\Sigma}}_{\boldsymbol{\beta}})$, $Q_h(h) = \text{Gamma}(\tilde{b}, 1/\tilde{g})$, $Q_\varphi(\varphi) = \text{Gamma}(\tilde{e}, 1/\tilde{f})$ and $Q_{\alpha_p}(\alpha_p) = \text{Gamma}(\tilde{c}_p, 1/\tilde{d}_p)$, we have

$$\tilde{a} = a_0 + \sum_{i=1}^{N}\langle L_i\rangle, \ \tilde{h} = \langle h\rangle + \sum_{i=1}^{N}\langle \ln(1+e^{\psi_i})\rangle \quad (41)$$
$$\tilde{\boldsymbol{\Sigma}} = (\langle\varphi\rangle\mathbf{I} + \tilde{\boldsymbol{\Omega}})^{-1}, \ \tilde{\boldsymbol{\mu}} = \tilde{\boldsymbol{\Sigma}}[(\boldsymbol{y} - \langle r\rangle)/2 + \langle\varphi\rangle\mathbf{X}\boldsymbol{\beta}] \quad (42)$$
$$\tilde{\boldsymbol{\Sigma}}_{\boldsymbol{\beta}} = (\langle\varphi\rangle\mathbf{X}^T\mathbf{X} + \langle\tilde{\mathbf{A}}\rangle)^{-1}, \ \tilde{\boldsymbol{\mu}}_{\boldsymbol{\beta}} = \langle\varphi\rangle\tilde{\boldsymbol{\Sigma}}_{\boldsymbol{\beta}}\mathbf{X}^T\langle\boldsymbol{\psi}\rangle \quad (43)$$
$$\tilde{b} = a_0 + b_0, \ \tilde{g} = \langle r\rangle + g_0, \ \tilde{e} = e_0 + N/2 \quad (44)$$
$$\tilde{f} = f_0 + \frac{\langle\boldsymbol{\psi}^T\boldsymbol{\psi}\rangle}{2} - \langle\boldsymbol{\psi}\rangle^T\mathbf{X}\langle\boldsymbol{\beta}\rangle + \frac{\text{tr}[\mathbf{X}\langle\boldsymbol{\beta}\boldsymbol{\beta}^T\rangle\mathbf{X}^T]}{2} \quad (45)$$
$$\tilde{c}_p = c_0 + 1/2, \ \tilde{d}_p = d_0 + \langle\beta_p^2\rangle/2 \quad (46)$$

where $\tilde{\boldsymbol{\Omega}} = \text{diag}(\langle\omega_1\rangle, \cdots, \langle\omega_N\rangle)$, $\tilde{\mathbf{A}} = \text{diag}(\tilde{\alpha}_0, \cdots, \tilde{\alpha}_P)$, $\text{tr}[\boldsymbol{\Sigma}]$ is the trace of $\boldsymbol{\Sigma}$, $\langle \ln r\rangle$ and $\langle L_i\rangle$ are calculated as in (29), $\langle r\rangle = \tilde{a}/\tilde{h}$, $\langle\boldsymbol{\psi}^T\boldsymbol{\psi}\rangle = \tilde{\boldsymbol{\mu}}^T\tilde{\boldsymbol{\mu}} + \text{tr}[\tilde{\boldsymbol{\Sigma}}]$, $\langle\boldsymbol{\beta}\rangle = \tilde{\boldsymbol{\mu}}_{\boldsymbol{\beta}}$, $\langle\boldsymbol{\beta}^T\boldsymbol{\beta}\rangle = \tilde{\boldsymbol{\mu}}_{\boldsymbol{\beta}}^T\tilde{\boldsymbol{\mu}}_{\boldsymbol{\beta}} + \text{tr}[\tilde{\boldsymbol{\Sigma}}_{\boldsymbol{\beta}}]$, $\langle\boldsymbol{\beta}\boldsymbol{\beta}^T\rangle = \tilde{\boldsymbol{\mu}}_{\boldsymbol{\beta}}\tilde{\boldsymbol{\mu}}_{\boldsymbol{\beta}}^T + \tilde{\boldsymbol{\Sigma}}_{\boldsymbol{\beta}}$, $\langle h\rangle = \tilde{b}/\tilde{g}$, $\langle\varphi\rangle = \tilde{e}/\tilde{f}$ and $\langle\alpha_p\rangle = \tilde{c}_p/\tilde{d}_p$. Although we do not have analytical forms for $\gamma_{i1}$ and $\gamma_{i2}$ in $Q_{\omega_i}(\omega_i)$, we can use (36) to calculate $\langle\omega_i\rangle$ as

$$\langle\omega_i\rangle = \mathbb{E}_{r,\psi_i}[\mathbb{E}[\omega_i|r,\psi_i,y_i]] = (y_i + \langle r\rangle)\left\langle\frac{\tanh(\psi_i/2)}{2\psi_i}\right\rangle \quad (47)$$

where the mean property of the PG distribution[1] (Polson & Scott, 2011) is applied. To calculate $\langle\ln(1+e^{\psi_i})\rangle$ in (41) and $\left\langle\frac{\tanh(\psi_i/2)}{2\psi_i}\right\rangle$ in (47), we use the Monte Carlo integration algorithm (Andrieu et al., 2003).

## 5. Example Results

### 5.1. Univariate Count Data Analysis

The inference of the NB dispersion parameter $r$ by itself plays an important role not only for the NB regression (Lawless, 1987; Winkelmann, 2008) but also for univariate count data analysis (Bliss & Fisher, 1953; Clark & Perry, 1989; Saha & Paul, 2005; Lloyd-Smith, 2007), and it also arises in some recently proposed latent variable models for count matrix factorization (Williamson et al., 2010; Zhou et al., 2012). Thus it is of interest to evaluate the proposed closed-form Gibbs sampling and VB inference for this parameter alone, before introducing the regression analysis part.

We consider a real dataset describing counts of red mites on apple leaves, given in Table 1 of Bliss & Fisher (1953). There were in total 172 adult female mites found in 150 randomly selected leaves, with a 0 count on 70 leaves, 1 on 38, 2 on 17, 3 on 10, 4 on 9, 5 on 3, 6 on 2 and 7 on 1. This dataset has a mean of 1.1467 and a variance of 2.2736, clearly overdispersed. We assume the counts are NB distributed and we intend to infer $r$ with a hierarchical model as

$$y_i \overset{iid}{\sim} \text{NB}(r, p), \ r \sim \text{Gamma}(a, 1/b), \ p \sim \text{Beta}(\alpha, \beta)$$

where $i = 1, \cdots, N$ and we set $a = b = \alpha = \beta = 0.01$. We consider 20,000 Gibbs sampling iterations, with the first 10,000 samples discarded and every fifth sample collected afterwards. As shown in Figure 1, the autocorrelation of Gibbs samples decreases quickly as the lag increases, and the VB lower bound converges quickly even starting from a bad initialization ($r$ is initialized two times the converged value).

The estimated posterior mean of $r$ is 1.0812 with Gibbs sampling and 0.9988 with VB. Compared to the method of moments estimator (MME), MLE, and maximum quasi-likelihood estimator (MQLE) (Clark

---
[1] There is a typo in B.2 Lemma 2 and other related equations of Polson & Scott (2011), where $\mathbb{E}(\omega) = \frac{a}{c}\tanh(\frac{c}{2})$ should be corrected as $\mathbb{E}(\omega) = \frac{a}{2c}\tanh(\frac{c}{2})$.

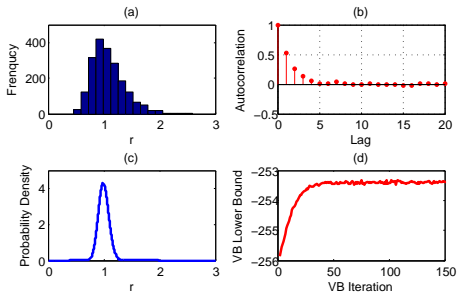

*Figure 1.* (a) The histogram and (b) autocorrelation of collected Gibbs samples of the NB dispersion parameter $r$. (c) The inferred probability density function of $r$ using VB. (d) The VB lower bound. Note that the calculated lower bound shows variations after convergence, which is expected since the Monte Carlo integration is used to calculate non-analytical expectation terms such as $\langle \ln \Gamma(r+y_i) \rangle$.

& Perry, 1989), which provides point estimates of 1.1667, 1.0246 and 0.9947[2], respectively, our algorithm is able to provide a full posterior distribution of $r$ and is convenient to incorporate prior information. The calculating details of the MME, MLE and MQLE, the closed-form Gibbs sampling and VB update equations, and the VB lower bound are all provided in the supplementary material, omitted here for brevity.

### 5.2. Regression Analysis of Counts

We test the full LGNB model on two real examples, with comparison to the Poisson, NB, lognormal-Poisson and inverse-Gaussian-Poisson (IG-Poisson) regression models. The NASCAR dataset[3], analyzed in Winner, consists of 151 NASCAR races during the 1975-1979 Seasons. The response variable is the number of lead changes in a race, and the covariates of a race include the number of laps, number of drivers and length of the track (in miles). The MotorIns dataset[4], analyzed in Dean et al. (1989), consists of Swedish third-party motor insurance claims in 1977. Included in the data are the total number of claims for automobiles insured in each of the 315 risk groups, defined by a combination of DISTANCE, BONUS, and MAKE factor levels. The number of insured automobile-years for each group is also given. As in Dean et al. (1989), a 19 dimensional covariate vector is constructed for each group to represent levels of the factors. To test goodness-of-fit, we use the Pearson residuals, a metric widely used in GLMs (McCullagh & Nelder, 1989), calculated as

$$E = \sum_{i=1}^{N} e_i^2, \quad e_i = \frac{y_i - \hat{\mu}_i}{\sqrt{\hat{\mu}_i(1 + \hat{\kappa}\hat{\mu}_i)}} \qquad (48)$$

where $\hat{\mu}$ and $\hat{\kappa}$ are the estimated mean and quasi-dispersion, respectively, whose calculations are described in detail in the supplementary material.

---

[2]The inverse dispersion parameter $\phi = 1/0.9947 = 1.005$ is mistakenly reported as the dispersion parameter $r$ in Clark & Perry (1989) at Line 15, Page 314.

[3]http://www.stat.ufl.edu/~winner/datasets.html
[4]http://www.statsci.org/data/general/motorins.txt

*Table 1.* The MLEs or posterior means of the lognormal variance parameter $\sigma^2$, NB dispersion parameter $r$, quasi-dispersion $\kappa$ and regression coefficients $\boldsymbol{\beta}$ for the Poisson, NB and LGNB regression models on the NASCAR dataset, using the MLE, VB or Gibbs sampling for parameter estimations.

| Model Parameters | Poisson (MLE) | NB (MLE) | LGNB (VB) | LGNB (Gibbs) |
|---|---|---|---|---|
| $\sigma^2$ | N/A | N/A | 0.1396 | 0.0289 |
| $r$ | N/A | 5.2484 | 18.5825 | 6.0420 |
| $\beta_0$ | -0.4903 | -0.5038 | -3.5271 | -2.1680 |
| $\beta_1$ (Laps) | 0.0021 | 0.0017 | 0.0015 | 0.0013 |
| $\beta_2$ (Drivers) | 0.0516 | 0.0597 | 0.0674 | 0.0643 |
| $\beta_3$ (TrkLen) | 0.6104 | 0.5153 | 0.4192 | 0.4200 |

*Table 2.* Test of goodness of fit with Pearson residuals.

| Models (Methods) | NASCAR | MotorIns |
|---|---|---|
| Poisson (MLE) | 655.6 | 485.6 |
| NB (MLE) | 138.3 | 316.5 |
| IG-Poisson (MLE) | N/A | 319.7 |
| LGNB ($r \equiv 1000$, Gibbs) | **117.8** | 296.7 |
| LGNB(VB) | 126.1 | **275.5** |
| LGNB(Gibbs) | 129.0 | 284.4 |

The MLEs for the Poisson and NB models are well-known and the update equations can be found in Winner; Winkelmann (2008). The MLE results for the IG-Poisson model on the MotorIns data were reported in Dean et al. (1989). For the lognormal-Poisson model, no standard MLE algorithms are available and we choose Metropolis-Hastings (M-H) algorithms for parameter estimation. We also consider a LGNB model under the special setting that $r = 1000$. As discussed in Section 3.1, this would lead to a model which is approximately the lognormal-Poisson model, yet with closed-form Gibbs sampling inference. We use both VB and Gibbs sampling for the LGNB model. We consider 20,000 Gibbs sampling iterations, with the first 10,000 samples discarded and every fifth sample collected afterwards. As described in the supplementary material, we sample from the PG distribution with a truncation level of 2000. We initialize $r$ as 100 and other parameters at random. Examining the samples in Gibbs sampling, we find that the autocorrelations of model parameters generally reduce to below 0.2 at the lag of 20, indicating fast mixing.

Shown in Table 1 are the MLEs or posterior means of key model parameters. Note that $\beta_0$ of the LGNB model differs considerably from that of the Poisson and NB models, which is expected since $\beta_0 + \sigma^2/2 + \ln r$ in the LGNB model plays about the same role as $\beta_0$ in the Poisson and NB models, as indicated in (16).

As shown in Tables 2, in terms of goodness of fit measured by Pearson residuals, the Poisson model performs the worst due to its unrealistic equal-dispersion assumption; the NB model, assuming a gamma distributed multiplicative random effect term, significantly improves the performances compared to the Poisson model; the proposed LGNB model, modeling extra-Poisson variations with both the gamma and lognormal distributions, clearly outperforms both the Poisson and NB models. Since for the lognormal-Poisson model with the M-H algorithm, we were not able to obtain comparable results even after carefully tuning the proposal distribution, we did not include it here for comparison. However, since the LGNB model reduces to the lognormal-Poisson model as $r \to \infty$, the results of the LGNB model with $r \equiv 1000$ would be able to indicate whether the lognormal distribution alone is appropriate to model the extra-Poisson variations. Despite the popularity of the NB model, which models extra-Poisson variations only with the gamma distribution, the results in Tables 2 suggest the benefits of incorporating the lognormal random effects. These observations also support the claim in (Winkelmann, 2008) that the lognormal-Poisson model should be reevaluated since it is appealing in theory and may fit the data better. Compared to the lognormal-Poisson model, the LGNB model has an additional advantage that its parameters can be estimated with VB inference, which is usually much faster than sampling based methods.

A clear advantage of the Bayesian inference over the MLE is that a full posterior distribution can be obtained, by utilizing the estimated posteriors of $\sigma^2$, $r$ and $\boldsymbol{\beta}$. For example, shown in Figure 2 are the estimated posterior distributions of the quasi-dispersion $\kappa$, represented with histograms. These histograms should be compared to $\kappa = 0$ in the Poisson model, and the NB model's MLEs of $\kappa = 0.1905$ and $\kappa = 0.0118$, for the NASCAR and MotorIns datasets, respectively. We can also find that VB generally tends to overemphasize the regions around the mode of its estimated posterior distribution and consequently places low densities on the tails, whereas Gibbs sampling is able to explore a wider region. This is intuitive since VB relies on the assumption that the posterior distribution can be approximated with the product of independent $Q$ functions, whereas Gibbs sampling only exploits conditional independence.

The estimated posteriors can also assist model interpretation. For example, based on $\tilde{\boldsymbol{\Sigma}}_{\boldsymbol{\beta}}$ in VB for the

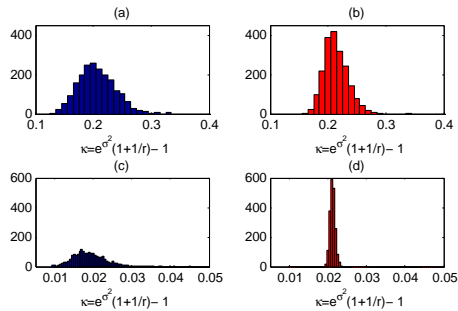

Figure 2. The histograms of the quasi-dispersion $\kappa = e^{\sigma^2}(1+1/r) - 1$ based on (a) the 2000 collected Gibbs samples for NASCAR, (b) the 2000 simulated samples using the VB $Q$ functions for NASCAR, (c) the 2000 collected Gibbs samples for MotorIns, and (d) the 2000 simulated samples using the VB $Q$ functions for MotorIns.

NASCAR dataset, we can calculate the correlation matrix for $(\beta_1, \beta_2, \beta_3)^T$ as

$$\begin{pmatrix} 1.0000 & -0.4824 & 0.8933 \\ -0.4824 & 1.0000 & -0.7171 \\ 0.8933 & -0.7171 & 1.0000 \end{pmatrix}$$

which is typically not provided in MLE. Since $\beta_1$ (Laps) and $\beta_3$ (TrkLen) are highly positively correlated, we expect the corresponding covariates to be highly negatively correlated. This is confirmed, as the correlation coefficient between the number of laps and the track length is found to be as small as $-0.9006$.

## 6. Conclusions

A lognormal and gamma mixed negative binomial (LGNB) regression model is proposed for regression analysis of overdispersed counts. Efficient closed-form Gibbs sampling and VB inference are both presented, by exploiting the compound Poisson representation and a Polya-Gamma distribution based data augmentation approach. Model properties are examined, with comparison to the Poisson, NB and lognormal-Poisson models. As the univariate lognormal-Poisson regression model can be easily generalized to regression analysis of correlated counts, in which the derivatives and Hessian matrixes of parameters are used to construct multivariate normal proposals in a Metropolis-Hastings algorithm (Chib et al., 1998; Chib & Winkelmann, 2001; Ma et al., 2008; Winkelmann, 2008), the proposed LGNB model can be conveniently modified for multivariate count regression, in which we may be able to derive closed-form Gibbs sampling and VB inference. As the log Gaussian process can be used to model the intensity of the Poisson process, whose inference remains a major challenge (Møller et al., 1998; Adams et al., 2009; Murray et al., 2010; Rao & Teh, 2011), we may link the log Gaussian process to the logit of the NB probability parameter, leading to a log Gaussian NB process with tractable closed-form Bayesian inference. Furthermore, the NB distribution is shown to be important for the factorization of a term-document count matrix (Williamson et al., 2010; Zhou et al., 2012), and the multinomial logit has been used to model correlated topics in topic modeling (Blei & Lafferty, 2005; Paisley et al., 2011). Applying the proposed lognormal-gamma-NB framework and the developed closed-form Bayesian inference to these diverse problems is currently under active investigation.

# Acknowledgements

The research reported here has been supported in part by DARPA under the MSEE program.